%% file: manuscript.tex
\documentclass[aps,prl,twocolumn,superscriptaddress,nofootinbib,floatfix]{revtex4-2}
\usepackage{graphicx}
\usepackage{amsmath,amssymb}
\usepackage{amsthm}
\newtheorem{theorem}{Theorem}
\newtheorem{corollary}{Corollary}
\newtheorem{proposition}{Proposition}
\usepackage{color}
\usepackage[colorlinks=true,linkcolor=blue,citecolor=blue,urlcolor=blue]{hyperref}

\begin{document}

\title{Commutator-Governed Energy Exchange in Single-Ancilla Coherent Feedback}

\author{Tan Jun Liang}
\email{junliang.tan@student.uq.edu.au}
\affiliation{School of Information Technology and Electrical Engineering, University of Queensland, St Lucia, QLD 4072, Australia}
% Add other authors here

\date{\today}

\begin{abstract}
A common route to ground-state preparation couples a system to one ancilla qubit,
applies a conditional feedback operation, and resets the ancilla. I ask what
physical quantity governs the energy such a cycle exchanges. The answer is a
commutator, not the correlation the ancilla acquires.

The energy change per cycle is exactly
$W = \langle\Psi_1|\,A - U^\dagger A U\,|\Psi_1\rangle$ with $A = I_A \otimes H$,
for every feedback strength and every sensing time, and it vanishes identically
whenever the feedback generator commutes with $H$. No function of the
system--ancilla correlation can therefore determine the work: on an explicit
one-parameter family the mutual information and the
logarithmic negativity are invariant to $2\times10^{-15}$ while the work varies
continuously and changes sign.

At fixed correlation the reachable work is a closed-form interval, and to leading
order in the feedback strength it is symmetric about zero, so an unfiltered
conditional kick heats as readily as it cools. That symmetry is exactly one
orthogonality condition on the two spectra, and the two sufficient conditions I
identify are its degenerate solutions, one holding automatically whenever the
post-sensing branch is pure. At that order no choice of generator repairs a
protocol that starts pure, and no choice of state repairs one with a symmetric
generator. Breaking both is necessary but not sufficient: across the grid I
searched, a finite kick's second-order bias raises the directional fraction only
to $0.015$. Directional
cooling instead requires the frequency-filtered jump operators of single-ancilla
Lindbladian preparation, which fall outside this Hermitian class.

Read in reverse the same commutator is constructive: it identifies, before any
circuit runs, variational parameter blocks whose energy gradient is identically
zero, removing a fifth of the circuits from an optimisation run.
\end{abstract}

\maketitle

% --- SECTION 1: INTRODUCTION ---
\section{Introduction}

Preparing ground states of many-body Hamiltonians is a central bottleneck for
quantum simulation, and is QMA-hard in general \cite{kitaev1999quantum}. One family
of protocols approaches it dissipatively: couple the system to a single ancilla
qubit, apply a conditional operation, reset the ancilla, and repeat. Ding, Chen and
Lin \cite{ding2024single} construct such a scheme whose stationary state is the
ground state of $H$, using one ancilla and a reset each cycle, and which succeeds
even when the initial state has zero overlap with the target. A closely related
structure arises in variational quantum algorithms, where an ancilla is used to
sense the system energy and drive a conditional update, and where trainability is
limited by the barren plateau phenomenon \cite{mcclean2018barren,
cerezo2021variational, ragone2024lie}.

Encoding variational parameters in a quantum register so that they evolve
coherently is not new. Verdon, Pye and Broughton introduced the Baqprop principle
and Quantum Dynamical Descent, in which error information is carried in the
relative phases of a wavefunction defined over the space of network parameters, and
optimization proceeds by coherent evolution of that register
\cite{verdon2018universal}. The protocol analysed here belongs to that family, and
no claim of priority for in-circuit parameter optimization is made.

The question examined below is narrower. Given a sense--actuate--reset cycle of
this kind, what physical quantity governs the energy change it produces?

There is a well-developed answer for a neighbouring class of protocols. Sagawa and
Ueda \cite{sagawa2008second} bound the work extractable under \emph{measurement}
feedback by the information the controller acquires, and daemonic ergotropy
\cite{francica2017, allahverdyan2004ergotropy} quantifies the gain a correlated
ancilla confers on measurement-assisted extraction. Manzano, Plastina and Zambrini
\cite{manzano2018optimal} sharpen the picture by showing that the \emph{type} of
correlation matters and not merely the magnitude of the mutual information.

Those results govern protocols containing a measurement. The cycle studied here
contains none: the feedback is a conditional unitary and the ancilla is reset
without ever being read. The question of what governs the energy exchange is
therefore open rather than settled, and it is not answered by transporting the
informational bound across the gap. It is answered by
Theorem~\ref{thm:identity}, which gives the energy change exactly, and the answer
is a commutator.

\begin{figure*}[htbp!]
    \centering
    \includegraphics[width=\textwidth]{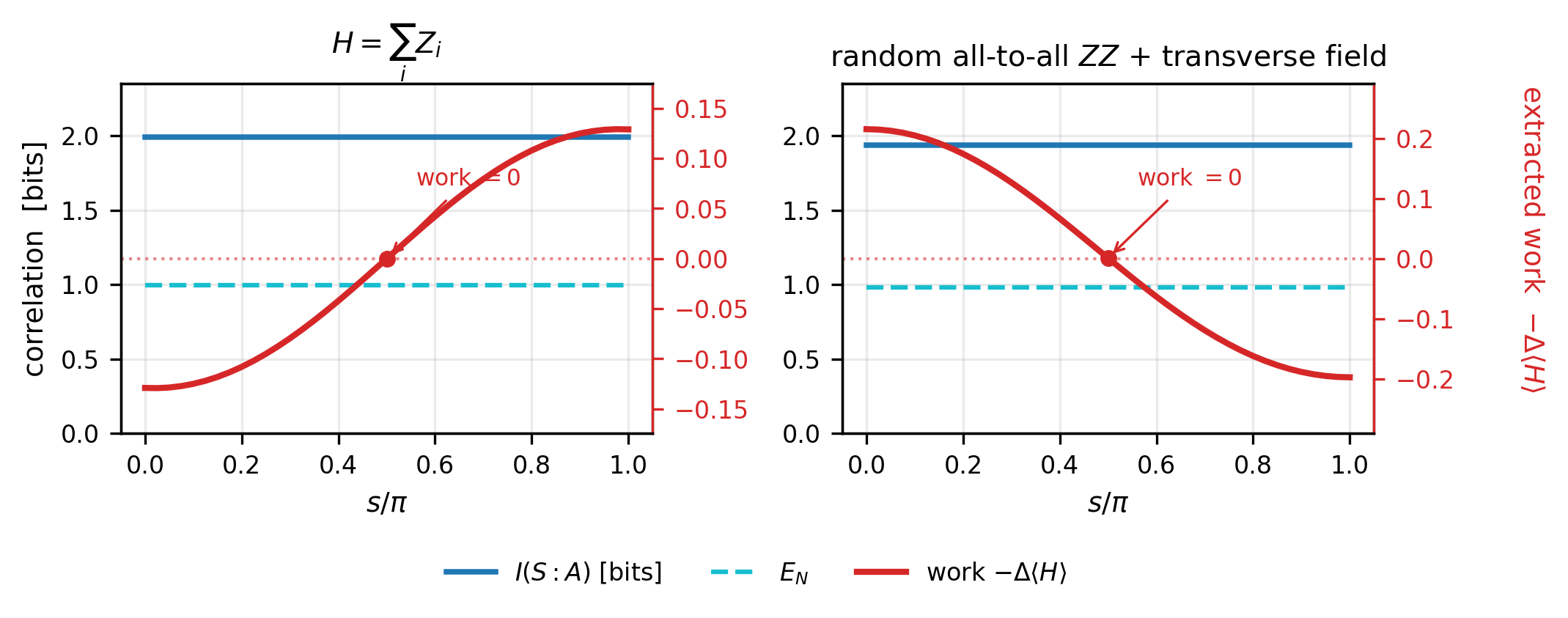}
    \caption{\textbf{No function of the system--ancilla correlation can determine
    the work.} Along the isospectral family $H(s) = V H V^\dagger$,
    $|\psi(s)\rangle = V|+\rangle^{\otimes n}$ with
    $V(s) = e^{-i s \sum_i Y_i / 2}$, the mutual information $I(S{:}A)$ (solid)
    and the logarithmic negativity $E_N$ (dashed) are constant while the extracted
    work $-\Delta\langle H\rangle$ (red) varies continuously and changes sign.
    $N=4$, $\theta = 0.2$, $\tau = 1.042$, exact statevector. Left: $H = \sum_i Z_i$,
    where the first-order work is proportional to $\cos s$ and vanishes at
    $s = \pi/2$. Right: a random all-to-all $ZZ$ instance with transverse fields,
    where the $s$-dependence has no closed form, showing the construction is a
    property of the protocol rather than of a special
    Hamiltonian. Over the full sweep $I(S{:}A)$ varies by $2.4\times10^{-15}$ and
    $4.2\times10^{-15}$ respectively and $E_N$ by $\leq 4.4\times10^{-15}$,
    machine precision, while the work spans $0.258$ and $0.413$. Any purported
    law $W = f(I(S{:}A))$ or $W = f(E_N)$ must assign a single value on each of
    these two lines. Theorem~\ref{thm:interval} generalises the picture: to first
    order in $\theta$, the reachable work at fixed correlation is the closed
    interval $[-W^*, +W^*]$.}
    \label{fig:isospectral}
\end{figure*}

The consequence is stronger than the qualitative statement of
Ref.~\cite{manzano2018optimal}. Not only does the magnitude of $I(S{:}A)$ fail to
fix the work: \emph{no} function of any correlation measure invariant under
system-local unitaries can, which is every standard one, because
Corollary~\ref{cor:zero} exhibits a system-local rotation that leaves all of them
invariant while moving the work. What that rotation does move is the alignment
between generator and Hamiltonian, which is what Eq.~\eqref{eq:firstorder}
supplies. Fig.~\ref{fig:isospectral} realises this
as a continuum on which the mutual information and the logarithmic negativity are
pinned to machine precision while the work varies continuously and changes sign.
The lesson generalises: an ancilla that becomes correlated with a system during a
controlled evolution is not thereby a demon, and a good linear fit across a
one-dimensional sweep does not establish that one quantity is a function of the
other.
Section~\ref{sec:mechanism} shows that the energy change is instead governed
exactly by the commutator of the feedback generator with the Hamiltonian, and
gives the full set of work values reachable while every system--ancilla correlation
is held fixed. Section~\ref{sec:cooling} places the protocol among single-ancilla
cooling schemes and identifies the component it is missing.

Figure~\ref{fig:isospectral} states the obstruction in its sharpest form. Rotating
the Hamiltonian and the initial state together by a system-local unitary $V(s)$,
while leaving the feedback generator fixed in the lab frame, sends the joint state
to $(I_A \otimes V(s))|\Psi_1\rangle$. A local unitary on one subsystem cannot
change either reduced spectrum, so every correlation measure is invariant along the
family, exactly and by construction rather than to within a tolerance. The work is not,
because the feedback did not rotate with them.

One preliminary observation motivates the setting. In a minimal four-qubit
circuit, two parameter qubits control two system qubits through conditional
rotations, the system evolves under $H$, and the encoding is inverted. For a
separable Hamiltonian $H = Z_0 + Z_1$ the accumulated phase factors as
$e^{-i(E_0 + E_1)t} = e^{-iE_0 t}e^{-iE_1 t}$, and the parameter register is left in
a product state at every evolution time:
\begin{equation}
    I(P_0 : P_1) \big|_{H = Z_0 + Z_1} = 0 ,
    \label{eq:nonlinearity}
\end{equation}
whereas the coupling term $H = Z_0 Z_1$ produces a non-separable phase and hence
correlation between the parameter qubits,
\begin{equation}
    I(P_0 : P_1) \big|_{H = Z_0 Z_1} = 1.74 \text{ bits at } t = 1 .
    \label{eq:nonlinearity_int}
\end{equation}
The second value is not generic in $t$: the two-body phase aliases into a product
of one-body phases at $t = k\pi/2$, where the correlation returns to zero for the
interacting Hamiltonian as well. The correct statement is therefore that a sum of
single-site terms leaves the parameter register uncorrelated at every evolution
time, while a coupling term does so only on a discrete set. The same aliasing
makes $I(S{:}A)$ quasi-periodic in the sensing time, so a scan over $\tau$ does
not parameterise a monotone increase in information.

\begin{figure}[htbp!]
    \centering
    \includegraphics[width=0.9\columnwidth]{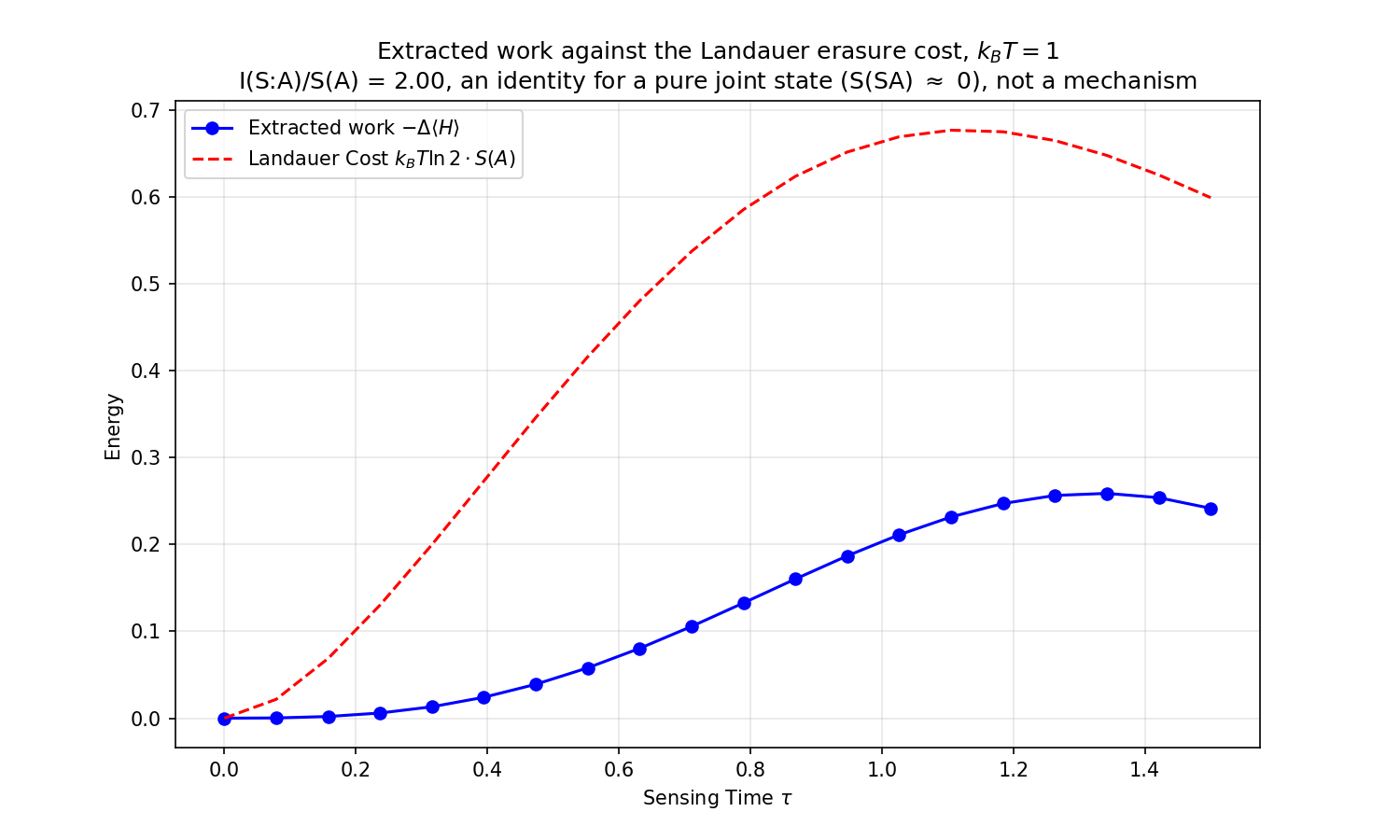}
    \caption{\textbf{Extracted work against the Landauer erasure cost.}
    Extracted work $-\Delta\langle H\rangle$ (blue) and $k_B T \ln 2 \cdot S(A)$
    (red dashed) as a function of sensing time $\tau$, at $N=4$ under exact
    statevector evolution, plotted at $k_B T = 1$. The couplings are
    dimensionless, so $k_B T$ is a modelling choice rather than a physical
    temperature; the curves cross at $k_B T^{*} = 0.406$
    [Eq.~\eqref{eq:landauerthreshold}], and the ordering shown holds only above
    it. The ratio $I(S{:}A)/S(A)$ is $2.00$ throughout, which is an identity for
    a pure joint state and is not evidence of a quantum mechanism.}
    \label{fig:landauer}
\end{figure}

What the Landauer comparison does show is retained in Fig.~\ref{fig:landauer}, and
it has to be stated conditionally. The Hamiltonian couplings here are
dimensionless, drawn from $U(-1,1)$ and $U(-0.5,0.5)$, so no physical temperature
is defined and the inequality $W \leq k_B T \ln 2 \cdot S(A)$ is true for large
enough $k_B T$ by construction. The content is the threshold at which it fails.
Writing $T_{\mathrm{req}}(\tau) = W(\tau) / [\ln 2\, S_A(\tau)]$, the claim that
the erasure cost covers the work at every sensing time is exactly the claim
$k_B T \geq \max_\tau T_{\mathrm{req}}(\tau)$, and that maximum is
\begin{equation}
  k_B T^{*} \;=\; 0.406 \quad\text{(attained at } \tau = 1.42\text{)},
  \label{eq:landauerthreshold}
\end{equation}
in units of the Hamiltonian couplings, at $N = 4$ and $\theta = 0.2$. The work
extracted per cycle therefore stays below the erasure cost for
$k_B T \geq 0.406$, with $2.5\times$ margin at the $k_B T = 1$ used in the figure,
and the ordering reverses below it.

The marginal $S(A)$ is the right entropy only for a reset that cannot access the
system the ancilla is correlated with. An erasure that may condition on it pays
the conditional entropy, which is smaller and can be negative
\cite{delrio2011thermodynamic}. Equation~\eqref{eq:landauerthreshold} therefore
bounds one reset strategy, not the thermodynamic cost of the protocol, and the
no-go itself is algebraic and carries no temperature.

\input{theorem_section}

\section{Relation to Single-Ancilla Cooling}
\label{sec:cooling}

The protocol analysed here (couple one ancilla to the system, apply a
conditional operation, reset the ancilla, repeat) is an instance of
single-ancilla dissipative state preparation. Ding, Chen and Lin
\cite{ding2024single} construct a Lindbladian of exactly this form, simulated with
one ancilla qubit and a reset each cycle, whose stationary state is the ground
state of $H$. Their algorithm prepares the ground state even when the initial
state has zero overlap with it, which is the principal limitation of phase
estimation.

The difference between that construction and the protocol studied here is a
frequency filter, and the filter is the entire cooling mechanism. Their jump
operator is
\begin{equation}
    K = \int_{-\infty}^{\infty} f(s)\, e^{iHs} A e^{-iHs}\, ds ,
\end{equation}
built from a Hermitian coupling operator $A$ and a filter satisfying
$\hat{f}(\omega) = 0$ for all $\omega \geq 0$. That condition forces
$\langle \psi_i | K | \psi_j \rangle = 0$ whenever $\lambda_i \geq \lambda_j$, so
energy increments are forbidden and the ground state is an exact fixed point.

The feedback studied here has no such filter: it is a fixed conditional rotation
applied at a single sensing time, with no weighting over the Heisenberg evolution
of the coupling operator. Theorem~\ref{thm:interval} shows the consequence
directly. To first order in $\theta$, the reachable work at fixed correlations is
the \emph{symmetric} interval $[-W^*, +W^*]$, because the spectrum of the feedback
generator is symmetric about zero and nothing breaks the sign degeneracy at that
order. An unfiltered conditional kick therefore heats as readily as it cools to
leading order, which accounts for every sign change reported above and for the
fold in Sec.~II.

Proposition~\ref{prop:twoconditions} names what would have to change, and it is
tempting to read that as a repair available within the unitary setting: break both
conditions and the interval tilts. It does tilt, and by almost nothing. Define the
directional fraction
\begin{equation}
  \mathcal{D} \;=\; \frac{W_{\mathrm{hi}} + W_{\mathrm{lo}}}
                         {W_{\mathrm{hi}} - W_{\mathrm{lo}}} \in [-1, 1],
  \label{eq:directional}
\end{equation}
zero for a symmetric interval and $\pm 1$ for a one-sided one, which is what
cooling requires. I sweep both conditions continuously at $N=3$, interpolating the
generator from $\sum_i X_i$ to a rank-one projector, mixing the branch by $1$ to
$12$ cycles, and making the Hamiltonian complex by a Dzyaloshinskii--Moriya term of
strength up to unity. The largest $|\mathcal{D}|$ anywhere on the grid is
$0.0147$. Even at a maximally asymmetric generator, with
$\mathrm{spec}(Y)$ defect $1.00$, the reachable interval remains symmetric to one
and a half percent. The tilt is also not monotone in mixing: it peaks near four
cycles and falls again by twelve, as the spectral defect of $M_{11}$ washes out.

The conclusion is therefore stronger than Proposition~\ref{prop:twoconditions}
alone. Breaking both conditions is necessary for any directionality, and on the
evidence here it is nowhere near sufficient for useful directionality. The
frequency filter is not one repair among several: within conditional unitary
feedback of this form, no amount of breaking these two symmetries buys a
preferred direction worth having.

\subsection{The filter works, and it works by leaving the class}

It is worth being precise about why the filter escapes, because the reason is not
the one Proposition~\ref{prop:twoconditions} would suggest. In the eigenbasis of
$H$ the filtered operator has entries
$K_{ij} = \hat{f}(\lambda_i - \lambda_j)\,A_{ij}$, so a filter supported on
$\omega < 0$ carries amplitude downward in energy only. \emph{That operator is not
Hermitian.} It does not satisfy condition~(A) or~(B) by some fortunate choice of
spectrum; it is not a Hermitian generator of a conditional unitary at all, and the
class the propositions describe does not contain it.

Built explicitly with a Gaussian filter at $N=3$,
$\|K - K^\dagger\|_F/\|K\|_F = 1.40$,
energy-lowering matrix elements of magnitude $1.66$ against energy-raising ones
suppressed to $3.9\times10^{-3}$. Run as a collision model with ancilla reset, it
cools monotonically from $-0.777$ to $-1.371$ over forty cycles, reaching
$94.5\%$ of the way to the ground state at $-1.406$. The same protocol with the
\emph{unfiltered} Hermitian coupling moves the energy by \emph{exactly zero} over
the same forty cycles.

\paragraph*{The price, and it is the point.} $K$ is built from $e^{iHs}$ over a
range of $s$, so a discretised filter costs one Hamiltonian evolution per sample,
\emph{every cycle}. Sweeping the number of samples:

\begin{center}
\begin{tabular}{c|c|r}
\hline
samples & rel.\ error in $K$ & progress to ground \\
\hline
$8$   & $1.13$   & $-55.6\%$ \\
$16$  & $0.73$   & $-76.3\%$ \\
$32$  & $0.0031$ & $+94.3\%$ \\
$64$  & $0.0000$ & $+94.5\%$ \\
\hline
\end{tabular}
\end{center}

Two things follow. Roughly $32$ distinct Hamiltonian evolutions per cycle are
needed at this size, and \emph{an under-resolved filter does not cool weakly, it
heats}: the sign of the effect inverts below the threshold. A partially
implemented filter is worse than none.

That count should not be read as the cost of a filtered repair in general. The
window used here is a Gaussian bandpass centred on a single negative frequency,
not the low-pass of Ref.~\cite{ding2024single}, which has support across the whole
negative spectrum. A narrow window is cheap \emph{because} it is narrow, and for
the same reason it drives fewer of the available transitions as the spectrum
widens: the cycles needed to reach $80\%$ of the way to the ground state grow from
$9$ at $N=3$ to beyond $400$ by $N=5$. Cheapness and failure here have the same
cause, so the sample count above bounds the cost of a filter at one size and says
nothing about the cost of one that works at scale.

The protocol studied here exists because phase estimation is too deep to run. Its
only repair puts Hamiltonian simulation back inside every cycle. That is not a
reason to abandon the approach, since the filtered construction demonstrably works
and Ref.~\cite{ding2024single} gives a rigorous version of it; but it does mean
the saving that motivated the unfiltered kick is not available, and a design that
keeps the shallow kick in order to avoid simulating $H$ is keeping the part that
provably cannot cool.

Two further observations bear on the use of this protocol as a benchmark.
First, there are two distinct ways a benchmark pairing can be degenerate, and
they call for different remedies. If the initial state is an eigenstate of $H$
the controlled evolution produces only a global phase, so $I(S{:}A) = 0$ and
$W = 0$ for every feedback generator and every sensing time; a single measurement
of $\Delta_H$ on the initial state detects this before any run, and a different
input repairs it. The isotropic Heisenberg chain paired with
$Y = \sum_i X_i$ fails for the stronger reason. That chain is $SU(2)$-symmetric
and therefore commutes with every total-spin operator, so $[H, Y] = 0$ and
Corollary~\ref{cor:zero} forces $W = 0$ for \emph{every} state, not only for
eigenstates. I verified this directly: with $|+\rangle^{\otimes n}$, which is
also an eigenstate, and with four Haar-random non-eigenstates, the exact work is
$\leq 1.6\times10^{-16}$ in all five cases at $N = 4$. Changing the input cannot
repair a commuting pairing; only a generator outside the symmetry algebra can.
Second, the
protocol's sensing step uses controlled Hamiltonian evolution, whereas
Ref.~\cite{ding2024single} requires only controlled-$A$ for a simple local
coupling operator; on gate cost the filtered construction is the cheaper of the
two.
\section{A Constructive Consequence}
\label{sec:constructive}

Corollary~\ref{cor:zero} is stated above as an obstruction: the commutator
$[H,Y]$ controls the energy exchange, and its vanishing forbids work in either
direction. The same statement is a \emph{predictor}, and in that direction it is
useful.

In a variational circuit whose parameters sit on single-qubit rotations, each
parameter $\theta_j$ has a generator $G_j$, and the energy gradient
$\partial_j \langle H \rangle$ is governed by the same commutator: it vanishes
identically when $[G_j, H_{\mathrm{eff}}] = 0$, with
$H_{\mathrm{eff}} = W^\dagger H W$ the Hamiltonian conjugated by the portion of
the circuit following $\theta_j$. The test is symbolic and costs no circuits.

This is a corollary rather than a parallel. Proposition~\ref{prop:general} shows
the feedback work and the variational gradient are the same expression
$\mathrm{Tr}(MG)$ evaluated at different observables, $A = I_A \otimes H$ in one
case and $A = H_{\mathrm{eff}}$ in the other
[Eq.~\eqref{eq:gradinstance}], so the vanishing condition transfers without a
separate argument.

That a vanishing commutator suppresses the gradient is not a new observation, and
the connection is not claimed as one. Hassan \emph{et al.}
\cite{hassan2026imaginarity} study the closely related question for the final
mixer angle in QAOA, bounding that gradient by the cost-basis imaginarity and
identifying imaginarity as a necessary condition for trainability; their bound is
sharper than a commutator criterion in the regime it covers, since it also
constrains gradients that are small without being zero, and it extends to noise
models. What the present formulation adds is scope and exactness. The condition
$[G_j, H_{\mathrm{eff}}] = 0$ is necessary \emph{and} sufficient for exact
vanishing, applies to any parameter anywhere in the circuit rather than to a final
mixer angle, holds for arbitrary $H$ rather than a diagonal cost with a real
mixer, and is decided symbolically before any circuit is run.

The exactness is what distinguishes it from a small-gradient heuristic. For a
diagonal Hamiltonian such as MaxCut or Ising,
a terminal layer of $R_Z$ rotations commutes with $H$ term by term, so that entire
parameter block has identically zero gradient at every point in parameter space.
I verified $\|\nabla_{\mathrm{block}} E\| = 0$ to machine precision on a
four-qubit MaxCut instance. Since $R_Z$ layers are half of the blocks in the
standard \texttt{EfficientSU2} ansatz, a quarter of its parameters are inert on
such problems, and no amount of sampling will reveal this to a
gradient-magnitude threshold: the quantity being thresholded is exactly zero, and
a shot-noise estimate of it is not.

Removing those blocks before optimisation begins is therefore free and safe. In
the optimiser used for the benchmarks reported here, the symbolic test reduces
the cost of a twenty-epoch run from $180$ circuits to $140$ with no change in the
energy reached. The commutator that forbids work extraction in the feedback
protocol and the commutator that forbids training in the variational circuit are
the same object, and both follow from Corollary~\ref{cor:zero} with no
approximation and no restriction on the feedback strength.

\section{Conclusion}

For the sense--actuate--reset cycle studied here, the energy exchanged per cycle
is fixed by a commutator and not by any correlation the ancilla acquires. The
identity is exact in the feedback strength and the sensing time, and it vanishes
identically when the feedback generator commutes with the Hamiltonian. To first
order in the feedback strength, the set of work values reachable while every
correlation measure is held fixed is a closed-form interval symmetric about zero.

That symmetry is the useful part, because it makes the limitation structural
rather than a matter of tuning. At that order, zero work is reachable at unchanged
correlation for every Hamiltonian and every state, and no choice of Hermitian
generator breaks the symmetry once the post-sensing branch is pure; the
obstruction survives any ancilla count, non-product couplings, and every cycle of
a repeated run.

The symmetry is a first-order statement and not an exact one, which is the limit
of what the theorem buys. At finite $\theta$ the second-order term supplies a
directional tilt, and breaking both sufficient conditions as hard as the setting
allows buys a directional fraction of only $0.0147$ --- small enough that the
conclusion drawn from the first-order result survives, but not zero, and the
distinction matters for anyone reading the interval as an exact constraint.

The escape is therefore not a better generator. It is a different kind of
operator. A frequency-filtered coupling is not Hermitian, so it lies outside the
class the propositions describe, and it works: built explicitly at $N=3$ it cools
to $94.5\%$ of the way to the ground state where the unfiltered kick moves the
energy by exactly zero. Its cost is one Hamiltonian evolution per filter sample
per cycle, and an under-resolved filter does not cool weakly. It heats, the sign
of the effect inverting below roughly $32$ samples.

Three limits should be read with the results. The quantitative figures rest on
dense simulation at $N \leq 7$, so nothing here speaks to hardware noise; the
theorems, being algebraic, do not depend on that. The closed form of the interval
endpoint rests on von Neumann's trace inequality, the one ingredient not
machine-checked.

The third limit is the one that bounds what may be concluded about cost. The
filter used here is a Gaussian \emph{bandpass}, centred on a single negative
frequency, whereas Ref.~\cite{ding2024single} specifies a low-pass with
$\hat{f}(\omega) = 0$ only for $\omega \geq 0$ and support across the whole
negative spectrum. A fixed narrow window needs few time samples \emph{because} it
is narrow, and it also drives a shrinking fraction of the available transitions as
the spectrum widens with $N$: cycles to reach $80\%$ of the way to the ground
state went $9$, $24$, then beyond $400$ at $N = 5$ and $6$. The flat sample count
and the failing convergence are therefore the same fact, and the flat count is not
evidence that a working filter is cheap at scale.

\paragraph*{Future directions.} Three questions follow, in increasing order of
how much they would change the picture. First, and most concretely: what does a
filter that covers the whole negative spectrum cost as $N$ grows? Such a filter
must resolve a spectral width that itself grows with $N$, so its sample count
need not stay flat the way a fixed window's does. A count growing linearly in $N$
would leave the repair viable; anything substantially worse and the fix inherits
the cost of Hamiltonian simulation proper, which is the expense the shallow kick
existed to avoid. Nothing measured here settles this, and it is the question that
decides whether the escape route is practical or merely formal. Second, whether
Corollary~\ref{cor:tracenorm} extends beyond
pure branches: the endpoint is exactly a commutator trace norm when the branch is
pure, and coincides there with the Wigner--Yanase skew information, but
$[\rho,A]$ and $[\sqrt{\rho},A]$ part company off purity and it is not settled
whether a skew-information bound survives.

Third, and most speculative, whether the obstruction here is related to the
expressivity bounds now being derived for walk-based optimisation, where the
dimension of the dynamical Lie algebra is polynomially bounded for broad problem
classes. Both are statements that structural simplicity in the generating
operators limits what a protocol can do, and the generators for which
condition~(A) holds, sums of single-qubit Paulis, are also those generating
small algebras. Whether that is a coincidence of the examples or a common
mechanism is open, and answering it would say more about this class of protocols
than anything measured here.

\section{Methods}

\subsection{Non-Linearity Test}
To establish the source of non-linear dynamics, I constructed a minimal 4-qubit circuit comparing interacting vs. non-interacting Hamiltonians. Two parameter qubits $(P_0, P_1)$ control rotations on two system qubits $(S_0, S_1)$:
\begin{enumerate}
    \item \textbf{Initialization:} Parameter qubits prepared in superposition $|+\rangle$.
    \item \textbf{W-Gate (Encode):} Controlled-$R_Y(\pi)$ maps $|1\rangle_P \to |1\rangle_S$.
    \item \textbf{Evolution:} Apply $e^{-iHt}$ with $t=1.0$.
    \item \textbf{Inverse W-Gate:} Disentangle system, transferring phase to parameters.
    \item \textbf{Measure:} Compute $I(P_0:P_1)$ from the parameter register.
\end{enumerate}
For $H = Z_0 + Z_1$, the phase factors as $e^{-i(E_0+E_1)t}$, yielding $I = 0$. For $H = Z_0 Z_1$, the coupled phase creates entanglement, yielding $I = 1.74$ bits. This demonstrates that non-linearity requires interaction terms.

\subsection{The sense--actuate--reset protocol}
\label{sec:protocol}
The protocol analysed throughout consists of three unitary phases applied to the joint state $\rho_{SA} = \rho_S \otimes |0\rangle\langle 0|_A$, followed by a reset of the ancilla:

\textbf{1. Sensing (Interaction):} The ancilla is prepared in superposition $|+\rangle_A$. The system and ancilla interact via a controlled-unitary evolution $U_{sense}(\tau) = |0\rangle\langle 0|_A \otimes I_S + |1\rangle\langle 1|_A \otimes e^{-i H_S \tau}$. This maps the local energy gradient into the relative phase of the ancilla, effectively realizing a short-time Phase Estimation routine or a weak measurement of the operator $H_S$.

\textbf{2. Correlation:} A Hadamard gate on the ancilla converts the phase information into population differences in the $Z$ basis. At this point I record the mutual information $I(S{:}A)$ and the logarithmic negativity $E_N$. No projective measurement is performed; the correlation remains coherent.

\textbf{3. Feedback (Actuation):} I apply a coherent feedback operation
$U_{kick} = CR_X(\theta_{gain})$, in which the system undergoes a rotation
conditioned on the ancilla state, with no projective measurement at any point.
The kick is sometimes motivated by analogy with a quantum walk coin, one branch
diffusing while the other traps. That analogy should not be read into the results:
Theorem~\ref{thm:interval} shows the reachable work is a symmetric interval to
first order in $\theta$, so an unfiltered kick has no preferred direction to
supply at that order, and the trapping half of the analogy is precisely what the
protocol lacks.

The reduced state $\rho_S$ evolves non-unitarily once the ancilla is reset (traced out) after the feedback step. The \emph{energy} change, however, involves no entropy export. Tracing out the ancilla and then evaluating a system observable is identical to evaluating $A = I_A \otimes H$ on the joint state, so $\Delta E$ is accounted for entirely by unitary redistribution of energy between the system and the ancilla inside the closed joint register, and is already fixed before any reset occurs. $\Delta E$ is therefore not paid for by entropy $S_{anc}$ exported to the environment at the reset: the reset changes the state, not the energy budget.

\subsection{Simulation Hyperparameters}
All numerical experiments in this version were performed with the parameters
listed in Table~\ref{tab:hyperparameters}, under exact statevector evolution
throughout.

\begin{table}[h]
\centering
\caption{\textbf{Simulation Hyperparameters.}}
\label{tab:hyperparameters}
\begin{tabular}{l|c|l}
\hline
\textbf{Parameter} & \textbf{Value} & \textbf{Description} \\
\hline
$N$ (System Size) & $4 - 7$ & Qubits in system \\
$J$ (Coupling) & $1.0$ / $\pm 1.0$ & Ord. / Chaotic \\
$\theta_{gain}$ (Kick) & $0.2$ rad & Feedback angle \\
$\tau$ (Sensing) & $0.0 - 1.5$ & Sensing duration \\
$L$ (Depth) & $1$ & Ansatz layers \\
Trials & $5$ & Per data point \\
Method & Statevector & exact, all $N$ \\
\hline
\end{tabular}
\end{table}

\subsection{Work and Information Definitions}
I define the thermodynamic quantities as follows:
\begin{itemize}
    \item \textbf{Extracted Work:} $W = E_{\text{before}} - E_{\text{after}} = \langle H \rangle_{\rho_S^{(0)}} - \langle H \rangle_{\rho_S^{(f)}}$, where $\rho_S^{(0)}$ is the reduced system state after sensing (before feedback) and $\rho_S^{(f)}$ is the state after feedback. Throughout, $W$ is the system energy decrease across the feedback step, not the thermodynamic work of a full cycle, which would also account for the sensing coupling and the ancilla reset. No result below depends on identifying the two.
    \item \textbf{Mutual Information:} $I(S:A) = S(\rho_S) + S(\rho_A) - S(\rho_{SA})$, computed in base 2 (bits) using von Neumann entropy.
\end{itemize}
The thermodynamic boundary is drawn to enclose only the quantum
information-processing degrees of freedom, so no macroscopic control overhead
enters any quantity reported here.

\textbf{A quantity deliberately not defined.} No efficiency of the form
$\eta = dW/dI$ is reported. Corollary~\ref{cor:zero} and
Fig.~\ref{fig:isospectral} show that $W$ is not a function of $I(S{:}A)$, so a
regression slope between the two measures the particular path taken through a
sweep rather than any property of the protocol.

\subsection{Statistical Reporting}
All results reported here were obtained by exact statevector simulation with no
sampling noise, so quantities such as $I(S{:}A)$, $E_N$ and $W$ carry no
statistical error and are reported as computed. Where a Hamiltonian family
contains randomness, results are averaged over five instances and the spread is
reported; the ordered and non-interacting families are deterministic by
construction, are run once, and no error estimate is available or meaningful for
them. A two-tailed Welch $t$-test is used only where both compared samples contain
more than one instance. No bootstrap confidence intervals, effect sizes or
multiple-comparison corrections are reported, because none are implemented in the
accompanying code.

\textbf{Numerical hazards.} Four defaults and habits in the simulation stack
produce plausible output rather than errors.
Qiskit's \texttt{PauliEvolutionGate} defaults to \texttt{LieTrotter(reps=1)},
which is far from converged for the Hamiltonians used here; moving to exact
evolution changes a reported coefficient of determination by a factor of three at
$N = 7$. \texttt{DensityMatrix.expectation\_value} does not check that the
observable and the state have the same dimension, and returns a number when they
do not. An acceptance gate on regression quality admits fits to numerical
noise: a case whose extracted work is $10^{-31}$ clears $R^2 \geq 0.60$.

Finally, and least visible of the four: two scripts asserted in their docstrings
that they built the same Hamiltonian from the same seed, and did not. One coupled
all pairs, the other nearest neighbours only; because the extra couplings advance
the same seeded random stream, the transverse fields differed as well. The
threshold in Eq.~\eqref{eq:landauerthreshold} was consequently computed on a chain
and quoted against a complete-graph figure, understating $k_B T^{*}$ by a factor
of $3.5$ and overstating the margin at $k_B T = 1$ as $8.5\times$ rather than
$2.5\times$. Nothing in either script was wrong in isolation, and a seeded
random Hamiltonian looks reproducible precisely when it is not being compared.
The check that catches this is to assert agreement of the constructed operator
across scripts rather than of the seed. All results in this version use exact
evolution, dimension-checked observables, and a single shared Hamiltonian
construction.
All results are exact statevector simulations, with no sampling and no noise
model. Entropic quantities such as $S$, $I(S{:}A)$ and $E_N$ are computed from
the state directly; estimating them on hardware would require tomography and is
not attempted here. Nothing in this work therefore speaks to performance under
device noise.

\textbf{Code Availability:} The simulation code for reproducing all numerical
experiments is available at
\url{https://github.com/poig/self-research/tree/main/Quantum_AI/QLTO/theory_test}.
Every claim-bearing script writes a committed log beside it, and the README maps
each result here to the script and log that produced it. The \texttt{superseded/}
and \texttt{supplementary/} directories hold withdrawn material and support
nothing claimed here.

\textbf{Relation to earlier versions.} Versions 1--2 of this preprint proposed a
constitutive relation $\Delta E \leq \eta(\mathcal{H},\mathcal{A}) \cdot I(S{:}A)$
and an efficiency transition governed by Dynamical Lie Algebra dimension.
Corollary~\ref{cor:zero} rules out the first. The supporting sweep was also not
injective: $I(S{:}A)$ is quasi-periodic in $\tau$ and turns over at $\tau = 1.18$
while $W$ continues to rise, placing two fitted points $0.002$ bits apart with
work differing by $17\%$. The efficiency transition does not survive
normalisation by $\|H\|$, and the fits defining $\eta$ used a first-order Trotter
approximation far from convergence. All numerical results here are regenerated
with exact time evolution. Those versions also described a classical feedback
baseline in which the ancilla is measured in the $Z$ basis and a classical
$R_X(\pm\theta_{\text{gain}})$ is applied on the outcome, and reported that it
extracts zero net work. No measurement of that baseline is presented here and the
claim is withdrawn as untested; it is in any case not needed, since
Theorem~\ref{thm:identity} makes no reference to whether the correlation is
quantum or classical.

\subsection{Acknowledgments}
\begin{acknowledgments}
The author used a large language model (Claude, Anthropic) to assist with drafting portions of this manuscript and implementing the simulation code. The research concept, theoretical framework, experimental design, and interpretation of results are solely the author's contribution. The author takes full intellectual responsibility for all scientific content presented herein.
\end{acknowledgments}

\bibliography{references}

\end{document}

%% file: theorem_section.tex
% =====================================================================
% Mechanism section for v3.  \input{theorem_section} after the Methods
% description of the sensing/feedback protocol.
%
% Numerical support:
%   theory_test/harmonized_sweep.py        Theorem 1, identity to 3e-15
%   theory_test/isospectral_family.py      Corollary 1, illustration
%   theory_test/achievable_work_theorem.py Theorem 2, all four checks
%   theory_test/exact_interval_asymmetry.py  Scope: A(theta) = 2.137 theta and
%                                            the theta* = 0.111 ergotropy cross
%   theory_test/second_order_directed_work.py  Prop. secondorder + Cor. parity:
%                                            W2/W1 = 2.1368 predicts the measured
%                                            slope; blind-cooling fraction table
%   theory_test/walk_symmetry_classification.py  (A)/(B) under reset, real vs
%                                            complex H, multi-cycle
% =====================================================================

\section{The Commutator Governs the Energy Change}
\label{sec:mechanism}

\subsection{Setting}

Write $\mathcal{H} = \mathcal{H}_A \otimes \mathcal{H}_S$ for one ancilla qubit
and $n$ system qubits, $H$ for the problem Hamiltonian, and
$A = I_A \otimes H$ for its embedding in the joint space. Let
$|\Psi_1\rangle$ denote the joint state after the sensing and locking steps of
Sec.~\ref{sec:protocol}, and let the feedback be the unitary
\begin{equation}
  U(\theta) = e^{-i (\theta/2) K}, \qquad K = P_1 \otimes Y ,
  \label{eq:feedback}
\end{equation}
where $P_1 = |1\rangle\langle 1|_A$ and $Y$ is a Hermitian generator on the
system. The protocol of Sec.~\ref{sec:protocol} takes $Y = \sum_i X_i$, realised
as one $CR_X(\theta)$ per system qubit; the results below hold for any Hermitian
$Y$.

\subsection{The identity}

\begin{theorem}[Exact energy change]
\label{thm:identity}
The energy change of the system across one feedback step is
\begin{equation}
  W \;=\; \langle \Psi_1 |\, A - U^\dagger(\theta)\, A\, U(\theta) \,|\Psi_1\rangle ,
  \label{eq:identity}
\end{equation}
exactly, for every $\theta$ and every sensing time $\tau$.
\end{theorem}

\begin{proof}
$W = \mathrm{Tr}(\rho_S H) - \mathrm{Tr}(\rho_S' H)$ with
$\rho_S = \mathrm{Tr}_A |\Psi_1\rangle\langle\Psi_1|$ and
$\rho_S' = \mathrm{Tr}_A\, U |\Psi_1\rangle\langle\Psi_1| U^\dagger$. For any
system operator $H$, $\mathrm{Tr}\big[(\mathrm{Tr}_A \sigma) H\big] =
\mathrm{Tr}[\sigma (I_A \otimes H)]$, so both terms may be evaluated on the joint
state, giving Eq.~\eqref{eq:identity}.
\end{proof}

Theorem~\ref{thm:identity}, Corollary~\ref{cor:zero}, the tensor structure that
carries the hypothesis of Corollary~\ref{cor:zero} to the controlled protocol, and
the symmetry half of Theorem~\ref{thm:interval} are machine-checked in Lean~4
against Mathlib, and every theorem the development exports depends on exactly the
three standard axioms
\texttt{propext}, \texttt{Classical.choice} and \texttt{Quot.sound}, with no
\texttt{sorry}. The formal statements are stronger than those proved below in two
respects: they are given for an arbitrary finite index type rather than for the
sizes simulated, and Eq.~\eqref{eq:identity} is formalised with no unitarity
hypothesis and no expansion in $\theta$. What is \emph{not} formalised is the
closed form of the endpoint $W^*$ in Theorem~\ref{thm:interval}, which rests on
von Neumann's trace inequality and is not available in Mathlib; nothing in the
development assumes it, and the symmetry of the interval, the part that
forbids directional cooling at first order in $\theta$, does not depend on it.

The algebra here is elementary, and deliberately so: Eq.~\eqref{eq:identity} is
the definition of an expectation value evaluated before and after a unitary,
lifted to the joint register. The content of the theorem is not the derivation but
what it excludes. It holds for every $\theta$ and every $\tau$ with no
small-angle expansion, so nothing about $W$ is left for a separate informational
account to supply; every subsequent result in this section is obtained by
analysing the right-hand side rather than by adding structure to it. Readers for
whom Eq.~\eqref{eq:identity} is obvious are invited to note that it is
nonetheless incompatible with any relation of the form $W = f(I(S{:}A))$, and
that the incompatibility is what Fig.~\ref{fig:isospectral} and
Theorem~\ref{thm:interval} make quantitative.

Two consequences are worth stating explicitly. First, $W$ is fixed before the
ancilla is reset: the reset alters the state but not the energy budget, and no
entropy is exported to pay for $W$. Second, $\mathrm{Tr}(\rho_S H)$ prior to
feedback is independent of $\tau$, since $\rho_S = \tfrac12(|\psi\rangle\langle\psi| +
e^{-iH\tau}|\psi\rangle\langle\psi|e^{iH\tau})$ and
$e^{iH\tau} H e^{-iH\tau} = H$; the pre-feedback energy is therefore not a
$\tau$-dependent baseline.

\subsection{Zero work at unchanged correlations}

\begin{corollary}[Exact vanishing]
\label{cor:zero}
If $[H, Y] = 0$ then $W = 0$ identically, for every $\theta$ and every $\tau$.
Moreover, for any $H$ and any $Y$ there exists a system-local unitary $V$ such
that the rotated configuration $(V H V^\dagger,\, V|\psi\rangle)$ has
$W = 0$ while every system--ancilla correlation measure is unchanged.
\end{corollary}

\begin{proof}
$[A, K] = (I_A \otimes H)(P_1 \otimes Y) - (P_1 \otimes Y)(I_A \otimes H)
 = P_1 \otimes [H, Y]$, which vanishes when $[H,Y] = 0$. Then $U$ commutes with
$A$, so $U^\dagger A U = A$ and Eq.~\eqref{eq:identity} gives $W = 0$ to all
orders in $\theta$.

For the second claim, rotating the Hamiltonian and the initial state together by
a system-local $V$ gives, for the post-sensing state,
$e^{-iVHV^\dagger \tau} V|\psi\rangle = V e^{-iH\tau}|\psi\rangle$, hence
\begin{equation}
  |\Psi_1(V)\rangle = (I_A \otimes V)\, |\Psi_1\rangle .
  \label{eq:localunitary}
\end{equation}
A unitary acting on one subsystem alone leaves both reduced spectra invariant, so
$I(S{:}A)$, $S(A)$, $S(S)$ and the logarithmic negativity are unchanged. The
feedback, however, is fixed in the laboratory frame, so the configuration is
equivalent to holding $H$ fixed and replacing $Y$ by $\tilde{Y} = V^\dagger Y V$.
Choosing $V$ so that $\tilde{Y}$ is diagonal in an eigenbasis of $H$ gives
$[H,\tilde{Y}] = 0$ and $W = 0$ by the first part.
\end{proof}

Corollary~\ref{cor:zero} carries no approximation and no restriction on $\theta$.
It follows that no function $W = f\big(I(S{:}A)\big)$ can exist for this protocol
class, and the same argument applies verbatim to any other correlation measure,
including the logarithmic negativity.

The obstruction belongs to a recognisable family. Commutation with a conserved
quantity limits what a controlled interaction can achieve in the
Wigner--Araki--Yanase setting, and Marvian \cite{marvian2022restrictions} shows
more generally that symmetry restricts which unitaries a composite system can
realise. Corollary~\ref{cor:zero} is the statement of that kind appropriate to
energy exchange under conditional feedback: what is conserved is $H$, what is
restricted is the reachable work, and the restriction is exact rather than
asymptotic. Verified numerically over a full sweep of
$V$ at $N = 4$: $I(S{:}A)$ and $E_N$ vary by $2\times10^{-15}$ and
$4\times10^{-15}$ while $W$ traverses $[-0.129, +0.129]$ and passes through zero
to $7\times10^{-17}$.

One related reading is worth closing here, since it invites the opposite
conclusion. For a pure joint state $S(SA) = 0$, so $I(S{:}A) = S(S) + S(A) =
2S(A)$ identically; an observed ratio $I/S(A) = 2.00$ is therefore a property of
purity and carries no information about the protocol.

\subsection{The achievable interval}

The first-order response admits a closed form. Expanding
Eq.~\eqref{eq:identity},
\begin{equation}
  W = \frac{\theta}{2}\,\big\langle \Psi_1 \big|\, i[A, K] \,\big| \Psi_1 \big\rangle
      + O(\theta^2)
    = \frac{\theta}{2}\, \mathrm{Tr}\!\left( M_{11}\, Y \right) + O(\theta^2),
  \label{eq:firstorder}
\end{equation}
where the second equality uses cyclicity together with $K = P_1 \otimes Y$, and
\begin{equation}
  M_{11} \;=\; \big\langle 1 \big|\, i\,[\,\rho,\, A\,] \,\big| 1 \big\rangle ,
  \qquad \rho = |\Psi_1\rangle\langle\Psi_1| ,
\end{equation}
is the Hermitian system operator obtained from the $|1\rangle\langle 1|$ ancilla
block. The generator enters only through $Y$, and under system-local rotation $Y$
moves over its isospectral orbit.

\begin{theorem}[First-order achievable set]
\label{thm:interval}
Let $\lambda^{\downarrow}$ and $\lambda^{\uparrow}$ denote eigenvalues in
decreasing and increasing order. Over all system-local unitaries $V$, and to
first order in $\theta$, the work achievable at fixed system--ancilla
correlations is the interval
\begin{equation}
  W \in \frac{\theta}{2}\left[\;
    \sum_k \lambda^{\downarrow}_k(M_{11})\,\lambda^{\uparrow}_k(Y) \;,\;\;
    \sum_k \lambda^{\downarrow}_k(M_{11})\,\lambda^{\downarrow}_k(Y)
  \;\right].
  \label{eq:interval}
\end{equation}
If $\mathrm{spec}(Y)$ is symmetric about zero the interval is
$[-W^{*}, +W^{*}]$ with
$W^{*} = \tfrac{\theta}{2}\sum_k \lambda^{\downarrow}_k(M_{11})\lambda^{\downarrow}_k(Y)$,
and both endpoints are attained by $V = U_Y U_M^\dagger$, where $U_Y$ and $U_M$
diagonalise $Y$ and $M_{11}$ with eigenvalues in matching order.
\end{theorem}

\begin{proof}
Immediate from von Neumann's trace inequality applied to
$\mathrm{Tr}(M_{11} V^\dagger Y V)$, whose extrema over the unitary orbit are the
similarly- and oppositely-ordered pairings of the two spectra
\cite{vonneumann1937}. Symmetry of $\mathrm{spec}(Y)$ makes the increasing list
the negation of the decreasing one, giving $-W^*$ for the lower endpoint.
\end{proof}

The optimisation technique is not new and is not claimed as such: extremising a
trace against an isospectral orbit by von Neumann's inequality, with the optimum
attained by matching eigenvalue orderings, is the standard derivation of
ergotropy \cite{allahverdyan2004ergotropy}. What Theorem~\ref{thm:interval} adds
is the object it is applied to. Ergotropy asks for the best unitary acting on the
\emph{state}; here the state and the Hamiltonian are rotated together and the
\emph{feedback generator} is what moves over the orbit, so the resulting interval
describes work reachable at \emph{fixed correlation} rather than work reachable at
fixed spectrum. That is what makes the endpoints comparable across the family of
Fig.~\ref{fig:isospectral}, and it is why the symmetry of the interval carries a
physical conclusion: the spectrum of $\sum_i X_i$ is symmetric about zero for
structural reasons, so the achievable set is too, for every $H$ and every state.

\subsection{The endpoint is a commutator norm}
\label{sec:monotone}

Equation~\eqref{eq:interval} is a pairing of two spectra, which is sharp but
bespoke. It collapses to a norm. H\"older's inequality applied to the pairing
gives $W^* \leq \tfrac{\theta}{2}\|Y\|_\infty \|M_{11}\|_1$ for any branch, and
on a pure branch the inequality is saturated:

\begin{corollary}[Endpoint as a trace norm]
\label{cor:tracenorm}
If the post-sensing branch is pure then
\begin{equation}
  W^{*} \;=\; \frac{\theta}{2}\,\|Y\|_\infty \, \big\| M_{11} \big\|_1 ,
  \qquad M_{11} = \langle 1 |\, i[\rho, A]\, | 1 \rangle .
  \label{eq:tracenorm}
\end{equation}
\end{corollary}

\begin{proof}
By Corollary~\ref{cor:purefixed} a pure branch gives $\mathrm{rank}\,M_{11} \leq 2$
with eigenvalues $\pm\Delta$, so the pairing has only two nonzero terms,
$\Delta\,\lambda^{\downarrow}_{\max}(Y) - \Delta\,\lambda^{\downarrow}_{\min}(Y)$.
When $\mathrm{spec}(Y)$ is symmetric this is $2\Delta\|Y\|_\infty$, and
$\|M_{11}\|_1 = 2\Delta$.
\end{proof}

Numerically the bound remains tight away from purity as well: at $N = 3$ and
$N = 4$ with branches mixed to purity $0.51$ and $0.37$, the ratio
$W^*/(\tfrac{\theta}{2}\|Y\|_\infty\|M_{11}\|_1)$ is $0.981$ and $0.905$.

This matters for placement rather than for content. Commutator norms of the form
$\|[\rho, A]\|$ are the family used to quantify coherence between eigenspaces of
$A$; the Wigner--Yanase skew information $S_A(\rho) = -\tfrac12
\mathrm{Tr}[A,\sqrt{\rho}]^2$ is the standard member, and it coincides with
$\mathrm{Var}(A)$ exactly on pure states, verified here to $10^{-9}$. So on the
pure branches for which the protocol is defined, the reachable work is fixed by
the same object that measures the state's asymmetry with respect to $A$, which is
the claim Sec.~\ref{sec:resource} makes qualitatively. Away from purity the two
part company, since $[\rho,A]$ and $[\sqrt{\rho},A]$ differ, and
Eq.~\eqref{eq:tracenorm} becomes an inequality rather than an identity; whether a
skew-information bound survives in that regime is not settled here.

\subsection{The same interval governs any observable}
\label{sec:general}

Neither proof above uses the tensor structure $A = I_A \otimes H$. Only
Hermiticity is required, so both statements are instances of one lemma.

\begin{proposition}[General first-order response]
\label{prop:general}
Let $\rho$ be a state, $A$ and $G$ Hermitian, and $U = e^{-i\theta G/2}$. Then
\begin{equation}
  \langle A\rangle_\rho - \langle A\rangle_{U\rho U^\dagger}
    = \frac{\theta}{2}\,\mathrm{Tr}\!\big(M G\big) + O(\theta^2),
  \qquad M := i[\rho, A],
  \label{eq:generalresponse}
\end{equation}
and over the isospectral orbit $G \mapsto V^\dagger G V$ the reachable value lies
in the closed interval given by von Neumann's inequality applied to
$\mathrm{spec}(M)$ and $\mathrm{spec}(G)$, symmetric about zero whenever
$\mathrm{spec}(G)$ is.
\end{proposition}

Theorem~\ref{thm:interval} is Eq.~\eqref{eq:generalresponse} at
$A = I_A \otimes H$ and $G = P_1 \otimes Y$, where the ancilla factor is inert and
$M$ reduces to the block $M_{11}$. The variational gradient is the same
expression with no ancilla at all: for
$|\psi(\theta)\rangle = W e^{-i\theta G_j/2} V|0\rangle$,
\begin{equation}
  \frac{\partial}{\partial\theta} \langle H \rangle
    = -\frac{1}{2}\,\mathrm{Tr}\!\big(M G_j\big),
  \qquad M = i[\,|\phi\rangle\langle\phi|,\, H_{\mathrm{eff}}\,],
  \label{eq:gradinstance}
\end{equation}
with $H_{\mathrm{eff}} = W^\dagger H W$ and
$|\phi\rangle = e^{-i\theta G_j/2}V|0\rangle$. The sign is the convention that
work counts an energy \emph{decrease} while a gradient counts an increase.
Verified on random circuits against an independently computed parameter-shift
gradient, agreeing to $4\times10^{-16}$; the lemma itself was checked against
exact finite response with the remainder confirmed $O(\theta^2)$, and the
interval against $300$ Haar-random $V$ per dimension with zero violations and
saturation to $10^{-17}$.

The consequence is that Sec.~\ref{sec:constructive} is not an analogy. A
variational block whose generator commutes with $H_{\mathrm{eff}}$ has
$\mathrm{Tr}(MG_j) = 0$ and hence an identically zero gradient, by exactly the
argument that gives $W = 0$ in Corollary~\ref{cor:zero}.

\begin{figure*}[htbp!]
    \centering
    \includegraphics[width=\textwidth]{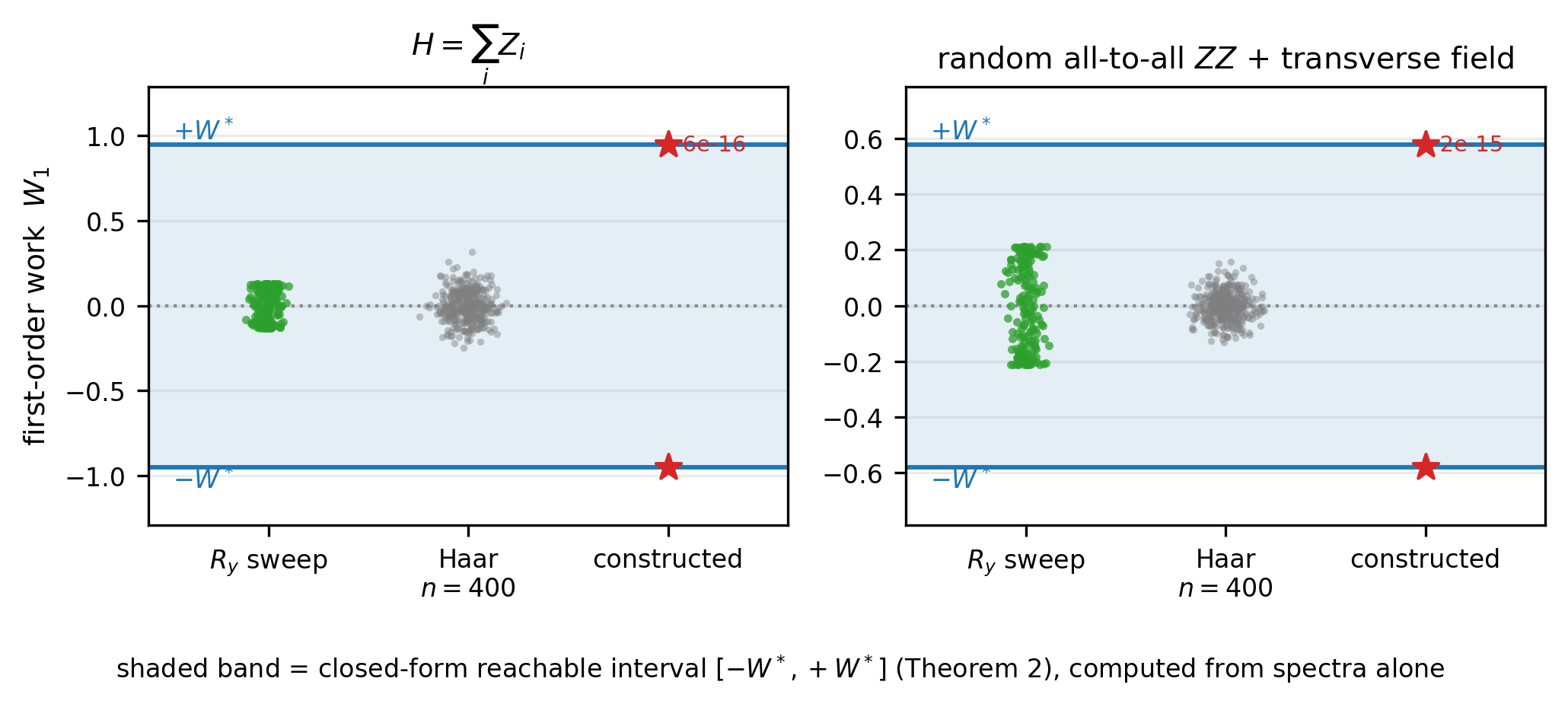}
    \caption{\textbf{The achievable set is predicted, not fitted.} The shaded band
    is Eq.~\eqref{eq:interval}, computed from the spectra of $M_{11}$ and $Y$
    alone, before any rotation is chosen. Green: the one-parameter $R_y$ family of
    Fig.~\ref{fig:isospectral}, which reaches only $36.6\%$ of the bound on the
    right-hand instance and so would badly understate the achievable range if
    taken as evidence on its own. Grey: $400$ Haar-random system-local $V$, every
    one inside the band, with \emph{zero violations}. Red stars: the explicit
    optimiser $V = U_Y U_M^\dagger$ of Theorem~\ref{thm:interval}, attaining both
    endpoints to $6\times10^{-16}$ and $2\times10^{-15}$. Along all of these the
    system--ancilla correlations are unchanged to $\leq 1.3\times10^{-15}$, so
    every point shown is reachable at fixed correlation. Zero lies at the centre
    of the band because the band is symmetric, which is
    Corollary~\ref{cor:zero}'s second part read off the figure: whatever the
    Hamiltonian and whatever the state, a frame exists in which this protocol does
    no work at all. $N=4$, $\theta=0.2$, $\tau=1.042$.}
    \label{fig:achievable}
\end{figure*}

\subsection{The resource is asymmetry, not correlation}
\label{sec:resource}

Equation~\eqref{eq:firstorder} identifies the resource explicitly. The system
operator $M_{11} = \langle 1 | i[\rho, A] | 1 \rangle$ is the $|1\rangle$ block of
$i[\rho, A]$, and since $A = I_A \otimes H$ acts as $H$ on each block it equals
$i[\rho_{11}, H]$ with $\rho_{11} = \langle 1 |\rho| 1 \rangle$. It therefore
vanishes exactly when the branch the feedback acts on commutes with $H$, that is,
when that branch is diagonal in the system's energy eigenbasis and carries no
coherence between energy levels. Commutation of the \emph{joint} state,
$[\rho, I_A \otimes H] = 0$, is sufficient for this but not necessary: it forces
every block to commute with $H$, whereas only the $|1\rangle$ block enters
Eq.~\eqref{eq:firstorder}. To first order in $\theta$, therefore,
\begin{equation}
  [\rho, I_A \otimes H] = 0 \;\Longrightarrow\; W = 0 ,
\end{equation}
for every feedback generator $Y$ and every rotation $V$. \emph{Energy coherence in
the branch the feedback acts on is what the protocol converts into work}, and
Eq.~\eqref{eq:interval} is the exact first-order conversion rate.

This places the protocol inside the resource theory of asymmetry, in which states
commuting with the Hamiltonian are free and coherence between energy eigenspaces
is the resource \cite{lostaglio2015coherence, bartlett2007reference}. It also
explains why an information-theoretic reading fails at the level of principle
rather than by accident. Lostaglio, Jennings and Rudolph showed that the
Szil\'ard-engine argument, work bounded by acquired information, does not
extend to quantum coherence, and that only \emph{relational} coherence in a
multipartite setting contributes to work
\cite{lostaglio2015beyondfree, korzekwa2016extraction}. Mutual information is not
an asymmetry monotone: the system-local rotation of
Corollary~\ref{cor:zero} leaves $I(S{:}A)$, $S(A)$, $S(S)$ and the logarithmic
negativity exactly invariant while moving $Y$ over its isospectral orbit, which
changes the alignment between $\mathrm{spec}(M_{11})$ and $\mathrm{spec}(Y)$ and
so sweeps $W$ across the whole interval. A correlation measure cannot see that
alignment, and the work depends on nothing else.

What Eqs.~\eqref{eq:identity} and \eqref{eq:interval} add to that literature is
exactness at the single-cycle level: an identity valid to all orders in the
feedback strength rather than a bound in an asymptotic or many-copy regime, and a
closed-form \emph{achievable set} rather than a one-sided inequality, obtained
because von Neumann's trace inequality is tight on the unitary orbit.

The reading also unifies two otherwise unrelated degeneracies of the protocol,
as the two sides of a single requirement: work needs a state that \emph{carries}
asymmetry and a generator that can \emph{couple} to it.
Corollary~\ref{cor:variance} supplies the first factor quantitatively, for a
pure post-sensing state the interval is proportional to $\Delta_H$, which is the
asymmetry monotone in this setting, the Wigner--Yanase skew information reducing
to the variance on pure states, so $\Delta_H = 0$, an energy eigenstate,
carries no resource and yields nothing for any generator. Corollary~\ref{cor:zero}
supplies the second: $[H,Y] = 0$ means the generator lies in the commutant of $H$
and cannot couple to the resource however much of it the state carries. The two
failures look alike in the data and are not alike in kind, and only the first is
repaired by choosing a different input.

\begin{corollary}[Pure-state form]
\label{cor:variance}
If the post-sensing state is pure, $M_{11}$ has rank at most two with eigenvalues
$\pm\Delta_H$, where $\Delta_H = \sqrt{\langle H^2\rangle - \langle H\rangle^2}$
is evaluated on the relevant marginal. Equation~\eqref{eq:interval} then reduces
to
\begin{equation}
  W \in \frac{\theta}{2}\,\Delta_H\big[\lambda_{\min}(Y) - \lambda_{\max}(Y),\;
                                        \lambda_{\max}(Y) - \lambda_{\min}(Y)\big].
  \label{eq:variancebound}
\end{equation}
\end{corollary}

\begin{proof}
$M_{11} \propto i(|\chi\rangle\langle\phi| - |\phi\rangle\langle\chi|)$ with
$|\phi\rangle = H|\chi\rangle$, hence is supported on
$\mathrm{span}\{|\chi\rangle, |\phi\rangle\}$. Writing $|\phi\rangle$ in an
orthonormal basis of that span and using $\langle\chi|H|\chi\rangle \in
\mathbb{R}$ gives eigenvalues $\pm\|(1 - |\chi\rangle\langle\chi|)H|\chi\rangle\|
= \pm\Delta_H$. Substituting into Eq.~\eqref{eq:interval} pairs $+\Delta_H$ with
$\lambda_{\max}(Y)$ and $-\Delta_H$ with $\lambda_{\min}(Y)$.
\end{proof}

The magnitude in Eq.~\eqref{eq:variancebound} is not new and should not be read
as such: it is Robertson's uncertainty relation $|\langle i[H,Y]\rangle| \leq
2\Delta_H \Delta_Y$ \cite{robertson1929} composed with Popoviciu's inequality
$\Delta_Y \leq (\lambda_{\max} - \lambda_{\min})/2$ \cite{popoviciu1935}, and the
optimising $Y$ saturates both simultaneously. The content of
Theorem~\ref{thm:interval} is not the size of the interval but the constraint
under which it is attained: Robertson bounds the gradient for a \emph{given}
generator on a \emph{given} state, and says nothing about moving the generator
over its isospectral orbit while every system--ancilla correlation measure is
held fixed. That invariance is supplied by Eq.~\eqref{eq:localunitary}, and it is
what makes Eq.~\eqref{eq:interval} a statement about correlations rather than
about magnitudes.

\paragraph{Relation to existing results.}
Neither tool is new and neither is claimed as a contribution. Equation
\eqref{eq:firstorder} is the first-order response of an expectation value to a
unitary generated by $K$, structurally the gradient used in optimal control
since Khaneja \emph{et al.} \cite{khaneja2005grape}, and the same object as the
parameter-shift rule. The ordered-pairing extremisation in
Theorem~\ref{thm:interval} is von Neumann's trace inequality, and is precisely
the computation underlying \emph{ergotropy}, the maximum work extractable from
a state by unitary means, with passive states as the minimisers
\cite{allahverdyan2004ergotropy}. What is established here is the achievable set
for \emph{this} protocol at fixed system--ancilla correlations, not the
inequality used to obtain it.

\paragraph{Scope.}
Theorem~\ref{thm:identity} and Corollary~\ref{cor:zero} are exact for all
$\theta$. Theorem~\ref{thm:interval} bounds the first-order term only. The exact
work at the optimising $V$ falls short of the first-order endpoint, by
$\approx 10\%$ at $\theta = 0.2$, $n = 4$, and the $O(\theta^2)$ term breaks
the exact symmetry of Eq.~\eqref{eq:interval}. No claim is made that the exact
achievable range is contained in the first-order interval.

Two quantitative limits on that statement matter for how
Sec.~\ref{sec:twocauses} should be read, since the symmetry of
Eq.~\eqref{eq:interval} is what makes the classification an obstruction.

First, \emph{the symmetry is exact only at $\theta = 0$}. Maximising and
minimising the exact $W$ of Eq.~\eqref{eq:identity} over $V \in U(2^n)$ directly,
rather than through the first-order pairing, gives a relative asymmetry
\begin{equation}
  \mathcal{A}(\theta) \;\equiv\;
  \frac{W_{\max} + W_{\min}}{W_{\max} - W_{\min}}
  \;=\; c\,\theta + O(\theta^2),
  \label{eq:asymscaling}
\end{equation}
with $c = 2.137$ measured at $n = 3$, $\tau = 0.35$: $\mathcal{A}/\theta$ is
$2.1369$, $2.1372$, $2.1384$ and $2.1433$ at $\theta = 0.0025$, $0.005$, $0.01$
and $0.02$. The constancy of that ratio over an eightfold range of $\theta$ is
what identifies the effect as the $O(\theta^2)$ term of
Eq.~\eqref{eq:firstorder} rather than a numerical artefact. Conditions (A) and
(B) therefore govern the \emph{linearised} protocol; a directed component exists
at any finite $\theta$ without either condition failing, and it is first order in
$\theta$ \emph{relative} to the interval width.

Second, \emph{the first-order interval is not everywhere achievable}. Only the
$|1\rangle$ branch is rotated and it carries weight $\tfrac12$, so
Eq.~\eqref{eq:identity} gives
$W = \tfrac12\big[\langle\phi|H|\phi\rangle -
\langle\phi|V^\dagger H V|\phi\rangle\big]$ and hence the exact bound
\begin{equation}
  W \in \tfrac12\big[\,\langle\phi|H|\phi\rangle - \lambda_{\max}(H),\;
                      \langle\phi|H|\phi\rangle - \lambda_{\min}(H)\,\big],
  \label{eq:ergobound}
\end{equation}
which is the ergotropy of the branch and is generally \emph{not} symmetric about
zero. At $n = 3$, $\tau = 0.35$ this floor is $-0.1028$ while the first-order
endpoint of Eq.~\eqref{eq:interval} grows as $0.9243\,\theta$, so the two cross
at $\theta^{*} = 0.111$: for $\theta > \theta^{*}$ the lower endpoint predicted
by Theorem~\ref{thm:interval} lies below anything the protocol can reach. At the
feedback strength used throughout this work, $\theta = 0.2$
(Table~\ref{tab:hyperparameters}), the predicted interval is $[-0.185, +0.185]$
against a reachable floor of $-0.103$, an overshoot of $80\%$ on that side.
Eq.~\eqref{eq:ergobound} is exact and involves no optimisation, so this is a
statement about the theorem's range of validity rather than about numerics.

\paragraph{Implementable frames and a two-level diagnostic.}
The optimising $V$ of Theorem~\ref{thm:interval} is an arbitrary element of
$U(2^n)$ and $M_{11}$ requires the joint state, so $W^{*}$ is a bound rather than
a prescription; the difference between $W^{*}$ and what a given circuit achieves
is not in general recoverable. For a feedback family $\mathcal{F}$ that the
hardware can realise, define $W^{*}_{\mathcal{F}}$ as the maximum of
Eq.~\eqref{eq:firstorder} over $\mathcal{F}$. The two gaps then separate:
$W^{*}_{\mathcal{F}}$ against the work actually obtained measures how well the
optimiser exploits the frame it has, while $W^{*}_{\mathcal{F}}$ against $W^{*}$
measures what the frame restriction costs. Only the first is actionable by
tuning.

\subsection{The second order is the directed part}
\label{sec:secondorder}

The scope note above records that the symmetry of Eq.~\eqref{eq:interval} is a
first-order statement. Carrying the expansion one term further identifies what
replaces it, and the structure of that term is what makes it useful.

\begin{proposition}[Second-order response]
\label{prop:secondorder}
\begin{equation}
  W \;=\; \frac{\theta}{2}\,\mathrm{Tr}(M_{11} Y)
        \;+\; \frac{\theta^{2}}{16}\,
              \big\langle \phi \big|\, [\,Y,[\,Y,H\,]\,] \,\big| \phi \big\rangle
        \;+\; O(\theta^{3}),
  \label{eq:secondorder}
\end{equation}
where $|\phi\rangle$ is the normalised $|1\rangle$ branch of $|\Psi_1\rangle$.
\end{proposition}

\begin{proof}
$U^\dagger A U = A + i(\theta/2)[K,A] - (\theta^2/8)[K,[K,A]] + O(\theta^3)$, so
by Theorem~\ref{thm:identity}
$W = (\theta/2)\langle i[A,K]\rangle + (\theta^2/8)\langle [K,[K,A]]\rangle
+ O(\theta^3)$. With $K = P_1 \otimes Y$ and $A = I_A \otimes H$ we have
$[K,A] = P_1 \otimes [Y,H]$, and since $P_1^2 = P_1$,
\begin{equation}
  [K,[K,A]] \;=\; P_1 \otimes \big[\,Y,[\,Y,H\,]\,\big].
  \label{eq:nested}
\end{equation}
Writing $|\Psi_1\rangle = (|0\rangle|\psi\rangle + |1\rangle|\phi\rangle)/\sqrt2$
gives $\langle \Psi_1 | P_1 \otimes M |\Psi_1\rangle = \tfrac12\langle\phi|M|\phi\rangle$
for any system operator $M$, and the first term is Eq.~\eqref{eq:firstorder}.
\end{proof}

\begin{corollary}[Parity, and the size of the tilt]
\label{cor:parity}
Under the reversal of the von Neumann pairing that maps one endpoint of
Eq.~\eqref{eq:interval} to the other, the first-order term changes sign and the
second-order term does not. Hence
\begin{equation}
  \mathcal{A}(\theta) \;=\;
    \frac{W_{\max}+W_{\min}}{W_{\max}-W_{\min}}
  \;=\; \frac{W_2}{W_1}\,\theta \;+\; O(\theta^{2}),
  \label{eq:paritytilt}
\end{equation}
with $W_1, W_2$ the coefficients of Eq.~\eqref{eq:secondorder}.
\end{corollary}

\begin{proof}
When $\mathrm{spec}(Y)$ is symmetric about zero the reversed pairing is realised
by $\tilde{Y} \mapsto -\tilde{Y}$. Then $\mathrm{Tr}(M_{11}\tilde{Y})$ changes
sign while $[\,-\tilde{Y},[-\tilde{Y},H]\,] = [\,\tilde{Y},[\tilde{Y},H]\,]$ is
unchanged, being quadratic in the generator. Writing
$W_{\max} = W_1\theta + W_2\theta^2$ and $W_{\min} = -W_1\theta + W_2\theta^2$
and forming the ratio gives Eq.~\eqref{eq:paritytilt}.
\end{proof}

At $n=3$, $\tau = 0.35$ this predicts $W_2/W_1 = 1.974944/0.924262 = 2.1368$
against the measured $2.1369$, $2.1372$, $2.1384$ of
Eq.~\eqref{eq:asymscaling}, agreement to four significant figures, from the
operators alone and with nothing fitted. The residual of
Eq.~\eqref{eq:secondorder} against the exact $W$ scales as $\theta^3$ over a
sixteenfold range, confirming that no term has been dropped.

\paragraph{Why this is the operationally relevant term.}
$\mathrm{Tr}(M_{11}Y)$ requires $M_{11}$, hence the joint state, which is why
$W^{*}$ is a bound rather than a prescription. The second-order term requires no
such alignment. In an eigenbasis of $Y$,
\begin{equation}
  [\,Y,[\,Y,H\,]\,] \;=\; \sum_{k,l} (y_k - y_l)^2\, H_{kl}\, |k\rangle\langle l| ,
  \label{eq:adsquared}
\end{equation}
a reweighting of $H$ by the squared spectral gaps of $Y$, manifestly independent
of how $\mathrm{spec}(Y)$ is \emph{ordered} against $\mathrm{spec}(M_{11})$.
Sampling $200$ system-local $V$ over the isospectral orbit, $W_1$ has mean
$+0.0004$ with range $[-0.313,+0.307]$ and changes sign, while $W_2$ has mean
$+0.691$ with range $[+0.086,+2.357]$ and does not. A generator chosen without
knowledge of the joint state therefore yields zero expected work at first order
and strictly positive work at second order.

\paragraph{A design rule, and the regime it requires.}
Because the two orders scale differently, a \emph{blind} protocol repeating a
fixed generator is directed only once $\theta$ exceeds $|W_1|/W_2$. Over $12$
cycles at $n=3$, with $24$ generators drawn at random from the isospectral orbit,
the fraction that cool is
\[
  \begin{array}{lcccccc}
  \theta & 0.05 & 0.10 & 0.20 & 0.30 & 0.40 & 0.60\\
  \text{cooling} & 54\% & 71\% & 83\% & 92\% & 100\% & 100\%
  \end{array}
\]
rising from a coin flip to certainty as $\theta$ crosses
$\max_V |W_1|/W_2 \approx 0.4$. The mean extracted energy over those generators
simultaneously approaches the aligned protocol's, from $11\%$ of it at
$\theta = 0.05$ to $53\%$ at $\theta = 0.6$. The feedback strength is therefore a
design variable rather than a small parameter, and the prescription runs opposite
to the weak-kick instinct: a protocol that cannot estimate $M_{11}$ should be
driven hard, where the sign-definite term dominates, rather than gently, where it
is dominated by a first-order term whose sign it does not control.
Eqs.~\eqref{eq:adsquared} and \eqref{eq:paritytilt} make this constructive:
choose $Y$ whose spectral gaps are large where $H$ carries off-diagonal weight,
and take $\theta$ above the crossover.

\subsection{Two independent causes of the symmetry}
\label{sec:twocauses}

Section~\ref{sec:cooling} reads the symmetric interval as a defect of the
unfiltered kick, to be repaired by a better feedback generator. That is only half
right, and the other half explains why the obvious repairs do not work.

\begin{proposition}[Exact criterion for the symmetry]
\label{prop:criterion}
Write $a = \lambda^{\downarrow}(M_{11})$, and let $b$ be the vector with entries
$b_k = \lambda^{\downarrow}_k(Y) + \lambda^{\uparrow}_k(Y)$. Then
\begin{equation}
  W_{\mathrm{hi}} + W_{\mathrm{lo}} \;=\; \frac{\theta}{2}\,\langle a,\, b\rangle ,
  \label{eq:criterion}
\end{equation}
so the interval of Eq.~\eqref{eq:interval} is symmetric about zero \emph{if and
only if} $\langle a, b\rangle = 0$. Neither spectrum is assumed to carry any
structure.
\end{proposition}

\begin{proof}
By Eq.~\eqref{eq:interval}, $W_{\mathrm{hi}} = \tfrac{\theta}{2}\sum_k a_k
\lambda^{\downarrow}_k(Y)$ and $W_{\mathrm{lo}} = \tfrac{\theta}{2}\sum_k a_k
\lambda^{\uparrow}_k(Y)$. Adding these and collecting the coefficient of each
$a_k$ gives Eq.~\eqref{eq:criterion}; symmetry is $W_{\mathrm{lo}} =
-W_{\mathrm{hi}}$, that is, the vanishing of the sum.
\end{proof}

Because $\lambda^{\uparrow}_k(Y) = \lambda^{\downarrow}_{n+1-k}(Y)$, the vector
$b$ is invariant under reversal of the index order. That one observation collapses
the two conditions below into the two degenerate ways an inner product vanishes.

\begin{proposition}[Two sufficient conditions]
\label{prop:twoconditions}
The interval of Eq.~\eqref{eq:interval} is symmetric about zero if \emph{either}
\begin{enumerate}
\item[(A)] $\mathrm{spec}(Y)$ is symmetric about zero, or
\item[(B)] $\mathrm{spec}(M_{11})$ is symmetric about zero.
\end{enumerate}
Each is sufficient on its own, for arbitrary choice of the other operator.
\end{proposition}

\begin{proof}
Both follow from Eq.~\eqref{eq:criterion}. Under (A),
$\lambda^{\downarrow}_{n+1-k}(Y) = -\lambda^{\downarrow}_k(Y)$, so $b = 0$ and the
inner product vanishes for every $M_{11}$. Under (B), $a_{n+1-k} = -a_k$ while
$b_{n+1-k} = b_k$ always, so pairing $k$ with $n+1-k$ cancels the terms, for every
$Y$.
\end{proof}

Conditions (A) and (B) are thus the two degenerate solutions of an orthogonality:
the first annihilates $b$, the second gives $a$ and $b$ opposite parity under
reversal. They are not exhaustive. Solving for the smallest eigenvalue of $M_{11}$
at generic $Y$ gives $\langle a, b\rangle = 0$ with neither spectrum symmetric in
$285$ of $500$ draws. Breaking both conditions is therefore necessary for an
asymmetric interval and not sufficient, which is how the small directional
fractions of Sec.~\ref{sec:cooling} should be read. The pure-branch obstruction is
case (B) and is untouched.

\paragraph*{Scope of the obstruction.} Nothing in
Proposition~\ref{prop:twoconditions} uses the single-ancilla structure, the
product form $K = P_1 \otimes Y$, or the first cycle, and dropping all three
leaves the conclusion intact, at $N = 3$ with
$|W_{\mathrm{lo}} + W_{\mathrm{hi}}|$ reported:
\begin{itemize}
\item \emph{Ancilla count.} $k = 1, 2, 3$ control qubits with the sensing
  conditioned on the ancilla register: symmetric at every $k$, to $0.0\times10^{0}$.
\item \emph{Non-product coupling.} Replacing $K$ by a general Hermitian operator
  on the joint space, including one whose own spectrum is \emph{asymmetric}
  (defect $0.812$): still symmetric. Condition~(B) fires on its own, because a
  first-cycle branch is pure and Corollary~\ref{cor:purefixed} then forces
  $\mathrm{spec}(M_{11})$ symmetric whatever the generator is.
\item \emph{Repeated operation.} Eight cycles of sense--actuate--reset, with the
  branch purity falling from $1.00$ to $0.47$ and $\mathrm{spec}(M_{11})$
  acquiring a defect of up to $2\times10^{-2}$: symmetric at every cycle,
  because condition~(A) continues to hold for the fixed generator.
\end{itemize}
The obstruction is therefore a property of the class rather than of the particular
protocol, and either condition alone carries it, which is why the restrictions can
be lifted one at a time without the conclusion moving.

Condition (A) is the one stated in Theorem~\ref{thm:interval}. Condition (B) is
not independent of the protocol's structure, and this is what closes off the
generator as a repair.

\begin{corollary}[A pure branch cannot be repaired by the generator]
\label{cor:purefixed}
If the post-sensing $|1\rangle$ branch is pure then (B) holds automatically, so
the interval is symmetric for \emph{every} Hermitian generator $Y$.
\end{corollary}

\begin{proof}
By Corollary~\ref{cor:variance} a pure branch gives $\mathrm{rank}\,M_{11}\le 2$
with eigenvalues $\{+\Delta_H, -\Delta_H, 0,\dots,0\}$, which is symmetric about
zero. Apply Proposition~\ref{prop:twoconditions}(B).
\end{proof}

Conversely, the generators natural to this protocol, sums of single-qubit
Paulis, for which $\mathrm{spec}(\sum_i X_i) = \{n-2k : k = 0,\dots,n\}$, satisfy
(A), so for those no choice of \emph{state} produces an asymmetric interval
either. The two corollaries together mean that testing one repair at a time can
only ever return a symmetric interval: an asymmetric reachable set requires (A)
and (B) to fail \emph{simultaneously}.

\paragraph*{The obstruction does not depend on system size.} That spectrum is
symmetric about zero at \emph{every} $n$, since $n-2k$ and $n-2(n-k)$ differ only
in sign. Condition~(A) therefore holds for the protocol's own generator at all
sizes, with no simulation required; checked numerically to $n=11$ and analytically
beyond, against a symmetry defect of $1.15$--$1.85$ for generic Hermitian
operators of the same dimension.
The system sizes reported in the Methods bound the \emph{quantitative} figures
only, namely the directional fraction and the filter's cost, which need dense
simulation. They do not bound the no-go itself.

The sensing time is not a third lever.

\begin{proposition}[Sensing-time invariance]
\label{prop:conjugation}
With $U_\tau = e^{-iH\tau}$ and $\rho_0$ the pre-sensing state,
$M_{11}(\tau) = U_\tau M_{11}(0) U_\tau^\dagger$, so $\mathrm{spec}\,M_{11}$ is
independent of $\tau$.
\end{proposition}

\begin{proof}
The branch is $\tfrac12 U_\tau \rho_0 U_\tau^\dagger$, so $M_{11}(\tau) \propto
i[U_\tau\rho_0 U_\tau^\dagger, H]$. Since $[U_\tau, H]=0$,
$i(U_\tau\rho_0 U_\tau^\dagger H - H U_\tau\rho_0 U_\tau^\dagger)
= iU_\tau(\rho_0 H - H\rho_0)U_\tau^\dagger$. Unitary conjugation preserves the
spectrum.
\end{proof}

I verified the classification numerically at $N=3$ and $N=4$, and it accounts for
four separate attempts to break the symmetry, each of which failed because it
addressed only one condition. A rank-$r$ projector generator breaks (A) but was
applied to a pure branch, where (B) holds by Corollary~\ref{cor:purefixed};
$r = 1,\dots,15$ all returned the identical interval. Admixing the identity leaves
$M_{11}$ merely rescaled, since $[I,H]=0$, and the same argument disposes of a
thermal input, for which $[e^{-\beta H},H]=0$ exactly. Multi-cycle mixing under a
real $H$ keeps $\mathrm{spec}\,M_{11}$ symmetric, for independent random
Hermitian pairs the normalised asymmetry of $\mathrm{spec}\,i[\rho,H]$ is
$\approx 0.3$, but it collapses below $10^{-15}$ whenever $\rho$ and $H$ are
simultaneously real, which the protocol's own families are. Adding a
Dzyaloshinskii--Moriya term $X_iY_j$ does break (B), giving a measured spectral
asymmetry of $4.4\times10^{-2}$, but with $Y=\sum_i X_i$ condition (A) still held
and the interval remained symmetric. Only breaking both at once, a rank-two
projector generator on a mixed branch under a complex $H$, produced an
asymmetric interval, $[-0.01910, +0.01921]$.

This sharpens what a cooling protocol must supply. It is not enough to filter, nor
to start mixed, nor to choose a cleverer generator: the construction needs a
spectrally asymmetric generator \emph{and} a branch whose commutator with $H$ is
spectrally asymmetric, and neither ingredient is useful without the other. Read
this way the filtered coupling of Ref.~\cite{ding2024single} supplies both at
once, which is why it works where a fixed kick does not: the filter
$\hat f(\omega)=0$ for $\omega\ge0$ produces a jump operator with no
$\pm$-symmetric spectrum, breaking (A), while the reset that makes the scheme
Lindbladian rather than unitary supplies the mixedness that breaks (B).

The scope is one cycle from a pure input, where
Corollaries~\ref{cor:variance} and~\ref{cor:purefixed} are exact. Across repeated
cycles the branch is no longer pure and (B) becomes a property of the accumulated
state; I observe it continuing to hold under real Hamiltonians and to fail under
complex ones, but have no proof that the classification is exhaustive there.

%% file: references.bib
@article{cerezo2021variational,
  author  = {Cerezo, M. and Arrasmith, A. and Babbush, R. and others},
  journal = {Nature Reviews Physics},
  title   = {Variational quantum algorithms},
  volume  = {3},
  pages   = {625-644},
  year    = {2021},
  doi     = {10.1038/s42254-021-00348-9}
}

@article{mcclean2018barren,
  title     = {Barren plateaus in quantum neural network training landscapes},
  author    = {McClean, J.R. and Boixo, S. and Smelyanskiy, V.N. and others},
  journal   = {Nature Communications},
  volume    = {9},
  pages     = {4812},
  year      = {2018},
  publisher = {Nature Publishing Group}
}

@article{francica2017,
  title     = {Daemonic ergotropy: enhanced work extraction from quantum correlations},
  author    = {Francica, G. and Goold, J. and Plastina, F. and Maniscalco, S.},
  journal   = {npj Quantum Information},
  volume    = {3},
  number    = {1},
  pages     = {12},
  year      = {2017},
  publisher = {Nature Publishing Group}
}

@article{ragone2024lie,
  title     = {A Lie algebraic theory of barren plateaus for deep parameterized quantum circuits},
  author    = {Ragone, Michael and Bakalov, Bojko N and Sauvage, Frédéric and Kemper, Alexander F and Marrero, Carlos Ortiz and Larocca, Martin and Cerezo, M.},
  journal   = {Nature Communications},
  volume    = {15},
  pages     = {7172},
  year      = {2024},
  publisher = {Nature Publishing Group},
  doi       = {10.1038/s41467-024-49909-3}
}

@article{verdon2018universal,
  author  = {Verdon, Guillaume and Pye, Jason and Broughton, Michael},
  title   = {A Universal Training Algorithm for Quantum Deep Learning},
  journal = {arXiv preprint arXiv:1806.09729},
  year    = {2018},
  doi     = {10.48550/arXiv.1806.09729}
}

@article{kitaev1999quantum,
  author    = {Kitaev, A. Yu.},
  title     = {Quantum {NP} and the Local Hamiltonian Problem},
  journal   = {Russian Mathematical Surveys},
  year      = {1999},
  note      = {QMA-completeness of the local Hamiltonian problem; see also
               Kitaev, Shen and Vyalyi, \emph{Classical and Quantum Computation},
               AMS Graduate Studies in Mathematics 47 (2002)}
}

@article{khaneja2005grape,
  author  = {Khaneja, Navin and Reiss, Timo and Kehlet, Cindie and
             Schulte-Herbr{\"u}ggen, Thomas and Glaser, Steffen J.},
  title   = {Optimal control of coupled spin dynamics: design of {NMR} pulse
             sequences by gradient ascent algorithms},
  journal = {Journal of Magnetic Resonance},
  volume  = {172},
  number  = {2},
  pages   = {296--305},
  year    = {2005},
  doi     = {10.1016/j.jmr.2004.11.004}
}

@article{allahverdyan2004ergotropy,
  author  = {Allahverdyan, A. E. and Balian, R. and Nieuwenhuizen, Th. M.},
  title   = {Maximal work extraction from finite quantum systems},
  journal = {Europhysics Letters},
  volume  = {67},
  number  = {4},
  pages   = {565--571},
  year    = {2004},
  doi     = {10.1209/epl/i2004-10101-2}
}

@article{vonneumann1937,
  author  = {von Neumann, J.},
  title   = {Some matrix-inequalities and metrization of matric-space},
  journal = {Tomsk University Review},
  volume  = {1},
  pages   = {286--300},
  year    = {1937},
  note    = {Reprinted in Collected Works, Vol. IV, Pergamon (1962)}
}

@article{robertson1929,
  author  = {Robertson, H. P.},
  title   = {The Uncertainty Principle},
  journal = {Physical Review},
  volume  = {34},
  number  = {1},
  pages   = {163--164},
  year    = {1929},
  doi     = {10.1103/PhysRev.34.163}
}

@article{popoviciu1935,
  author  = {Popoviciu, T.},
  title   = {Sur les {\'e}quations alg{\'e}briques ayant toutes leurs racines r{\'e}elles},
  journal = {Mathematica (Cluj)},
  volume  = {9},
  pages   = {129--145},
  year    = {1935}
}

@article{ding2024single,
  author  = {Ding, Zhiyan and Chen, Chi-Fang and Lin, Lin},
  title   = {Single-ancilla ground state preparation via {L}indbladians},
  journal = {Physical Review Research},
  volume  = {6},
  number  = {3},
  pages   = {033147},
  year    = {2024},
  doi     = {10.1103/PhysRevResearch.6.033147}
}

@article{lostaglio2015coherence,
  title   = {Quantum Coherence, Time-Translation Symmetry, and Thermodynamics},
  author  = {Lostaglio, Matteo and Jennings, David and Rudolph, Terry},
  journal = {Phys. Rev. X},
  volume  = {5},
  pages   = {021001},
  year    = {2015},
  doi     = {10.1103/PhysRevX.5.021001}
}

@article{lostaglio2015beyondfree,
  title   = {Description of quantum coherence in thermodynamic processes requires constraints beyond free energy},
  author  = {Lostaglio, Matteo and Jennings, David and Rudolph, Terry},
  journal = {Nature Communications},
  volume  = {6},
  pages   = {6383},
  year    = {2015},
  doi     = {10.1038/ncomms7383}
}

@article{korzekwa2016extraction,
  title   = {The extraction of work from quantum coherence},
  author  = {Korzekwa, Kamil and Lostaglio, Matteo and Oppenheim, Jonathan and Jennings, David},
  journal = {New J. Phys.},
  volume  = {18},
  pages   = {023045},
  year    = {2016},
  doi     = {10.1088/1367-2630/18/2/023045}
}

@article{bartlett2007reference,
  title   = {Reference frames, superselection rules, and quantum information},
  author  = {Bartlett, Stephen D. and Rudolph, Terry and Spekkens, Robert W.},
  journal = {Rev. Mod. Phys.},
  volume  = {79},
  pages   = {555},
  year    = {2007},
  doi     = {10.1103/RevModPhys.79.555}
}

@article{sagawa2008second,
  title = {Second Law of Thermodynamics with Discrete Quantum Feedback Control},
  author = {Sagawa, Takahiro and Ueda, Masahito},
  journal = {Phys. Rev. Lett.},
  volume = {100},
  pages = {080403},
  year = {2008},
  doi = {10.1103/PhysRevLett.100.080403}
}

@article{manzano2018optimal,
  title = {Optimal Work Extraction and Thermodynamics of Quantum Measurements and Correlations},
  author = {Manzano, Gonzalo and Plastina, Francesco and Zambrini, Roberta},
  journal = {Phys. Rev. Lett.},
  volume = {121},
  pages = {120602},
  year = {2018},
  doi = {10.1103/PhysRevLett.121.120602}
}

@article{marvian2022restrictions,
  title = {Restrictions on realizable unitary operations imposed by symmetry and locality},
  author = {Marvian, Iman},
  journal = {Nature Physics},
  volume = {18},
  pages = {283--289},
  year = {2022}
}

@misc{hassan2026imaginarity,
  title = {Imaginarity as a necessary resource for trainability in {QAOA}},
  author = {Hassan, Syed Muhammad Ali and Blekos, Kostas and K{\"u}hn, Stefan and Kollas, Nikos and Jansen, Karl},
  year = {2026},
  eprint = {2608.05093},
  archivePrefix = {arXiv},
  primaryClass = {quant-ph}
}

@article{delrio2011thermodynamic,
  title = {The thermodynamic meaning of negative entropy},
  author = {del Rio, L{\'\i}dia and {\AA}berg, Johan and Renner, Renato and Dahlsten, Oscar and Vedral, Vlatko},
  journal = {Nature},
  volume = {474},
  number = {7349},
  pages = {61--63},
  year = {2011},
  doi = {10.1038/nature10123}
}
